\documentclass[a4paper,12pt]{article}
\usepackage{psfig}

\def\bc{\begin{center}}
\def\ec{\end{center}}

\def\be{\begin{equation}}
\def\ee{\end{equation}}
\def\beq{\begin{eqnarray}}
\def\eeq{\end{eqnarray}}
\def\bfig{\begin{figure}}
\def\efig{\end{figure}}
\def\bnum{\begin{enumerate}}
\def\enum{\end{enumerate}}

\begin{document}

\vspace{1cm}
\bc
{\Large\bf Mechanisms of the Vertical Secular Heating of a Stellar Disk}\\
\vspace{0.7cm}
{\bf N. Ya. Sotnikova (nsot@astro.spbu.ru)\\ 
     S. A. Rodionov (seger@astro.spbu.ru)}\\
\vspace{0.7cm}
{\it Astronomical Institute, St. Petersburg State University,\\
Universitetskii pr. 28, Petrodvorets, St. Petersburg, 198904 Russia}\\
\ec

\bc
{\bf Abstract}
\ec

\bc
\parbox{15cm}{
\small
We investigate the nonlinear growth stages of bending instability in stellar 
disks with exponential radial density profiles.We found that the unstable 
modes are global (the wavelengths are larger than the disk scale lengths) and 
that the instability saturation level is much higher than that following from 
a linear criterion. The instability saturation time scales are of the order of 
one billion years or more. For this reason, the bending instability can play 
an important role in the secular heating of a stellar disk in the $z$ 
direction. In an extensive series of numerical $N$-body simulations with a 
high spatial resolution, we were able to scan in detail the space of key 
parameters (the initial disk thickness $z_0$, the Toomre parameter $Q$, and 
the ratio of dark halo mass to disk mass $M_{\rm h} / M_{\rm d}$). We 
revealed three distinct mechanisms of disk heating in the $z$ direction: 
bending instability of the entire disk, bending instability of the bar, and 
heating on vertical inhomogeneities in the distribution of stellar matter.
}
\ec
\vspace{0.5cm}
\bc
{\bf STELLAR RELAXATION IN THE DISKS OF SPIRAL GALAXIES}
\ec

According to the theory of stellar evolution, the correlation between the 
spectral types of stars in the solar neighborhood and their kinematic 
parameters found more than half a century ago (Parenago 1950) is the reflection 
of another correlation between the stellar age and the random velocity 
dispersion (for a review of the currently available data, see, e.g., Fuchs 
et al. 2000). The latter correlation has the following pattern: the velocity 
dispersion for old stars is, on average, larger than that for young stars. 
All stars are born with a small spread in random velocities (about 
5--10 km s$_{-1}$. In this case, the increase in the velocity dispersion with 
time (dynamic heating) is the reflection of a random scattering of stars 
(continuous or occasional) by massive objects or large-scale density 
inhomogeneities.

\medskip
Two pioneering studies (Spitzer and Schwarzshild 1951, 1953), in which large 
gaseous structures discovered twenty years later as giant molecular clouds 
(GMCs) from CO observations were postulated as scattering objects, were 
based on this idea. Subsequently, other relaxation mechanisms, for example, 
interaction with close satellites, were also suggested. The latter mechanism 
relies on observational data, according to which the disks of galaxies in 
interacting systems are a factor of 1.5 or 2 thicker than the disks of 
isolated galaxies (Reshetnikov and Combes 1997). Its efficiency is confirmed 
by numerical simulations (Walker et al. 1999; Velazquez and White 1999).

\medskip
For isolated galaxies, apart from scattering by GMCs, the heating on 
inhomogeneities in the distribution of stellar matter that arise during the 
development of internal instabilities in the disk itself is commonly 
considered. The heating can be produced by transient spiral arms or a 
growing bar and by bending instability.

\medskip
In suggesting a particular dynamic heating mechanism, the following two 
well-established observational facts should be borne in mind.

\medskip
(1) The stellar velocity dispersion depends on time as 
$\sigma_{\rm tot} \propto t^{\alpha}$, where $\sigma_{\rm tot}$ is the 
three-dimensional velocity dispersion and $\alpha \approx 0.33 - 0.5$ 
(Fuchs et al. 2000; Binney et al. 2000).

\medskip
(2) The ratio of the vertical and radial stellar velocity dispersions is 
$\sigma_z / \sigma_R < 1$.

\medskip
The second fact requires an explanation. It is well known that in an 
equilibrium axisymmetric galaxy in the absence of the third integral of 
motion, $\sigma_z$ and $\sigma_R$ must be equal at a given point of the 
disk (see, e.g., Saslaw 1989). Observational data indicate that this is not 
the case. The $\sigma_z / \sigma_R$ ratio for stars in the solar neighborhood 
does not depend on their spectral type (and, hence, on their age) and is 
$\sigma_z / \sigma_R = 0.53 \pm 0.07$ (see, e.g., Dehnen and Binney 1998). 
Until recently, no similar data have been available for external galaxies. 
At present, this ratio is known for two more nearby galaxies\footnote{In both 
cases, we give galaxy-averaged data.}:  
$\sigma_z / \sigma_R = 0.70 \pm 0.19$ for NGC 488 (morphological type Sb) 
(Gerssen et al. 1997) and 
$\sigma_z / \sigma_R =0.85 \pm 0.1$ for NGC 2985 (morphological type Sab) 
(Gerssen et al. 2000); i.e., it increases as one passes to earlier-type 
galaxies. 

\medskip
Spitzer and Schwarzshild (1951, 1953) showed that when stars are scattered 
by GMCs, the stellar velocity dispersion increases as 
$\sigma_{\rm tot}~\propto~t^{1/3}$; i.e., the exponent corresponds to the 
lower limit of this quantity, which is still in agreement with observational 
data. If, however, we take into account the fact that the GMCs are confined 
in a thin layer and the stars lie outside this layer most of the time, then 
the efficiency of the relaxation mechanism associated with GMCs becomes even 
lower: $\sigma_{\rm tot} \propto t^{0.25}$ (Lacey 1984). At the same time, 
GMCs are capable of effectively converting the energy of random motions in 
the disk plane into the energy of random motions in the perpendicular 
direction. Lacey (1984) found that when stars are scattered by GMCs, a 
ratio $\sigma_z / \sigma_R \approx 0.8$ is rapidly established. This ratio 
is much larger than the value observed for the solar neighborhood, although 
it is close to that obtained for the galaxy NGC 2985. However, no CO-line 
emission was detected for this galaxy (see, e.g., Merrifield et al. 2000).

\medskip
As regards the spiral arms, numerical simulations (the first numerical results 
were obtained by Sellwood and Carlberg 1984) and analytical calculations 
(Binney and Lacey 1988; Jenkins and Binney 1990; Jenkins 1992) show good 
agreement between the theoretical and observational age~--~velocity dispersion 
relations. However, the following should be remembered: in the above 
theoretical studies, the spiral arms were assumed to effectively scatter 
stars only in the disk plane. Additional mechanisms are commonly invoked 
to convert part of the energy in the disk plane into the energy of vertical 
motions. The combined effect from spiral arms and GMCs yields a result that 
is in good agreement with observational data (Binney and Lacey 1988), but 
this does not rule out other vertical disk heating mechanisms. For isolated 
galaxies, bending instability can serve as such a mechanism. In addition, as 
we show below, spiral arms are capable of producing density inhomogeneities 
in the $z$ direction. The interaction of stars with these density 
enhancements also result in an increase of $\sigma_{\rm z}$.

\medskip
Here, we study in detail the nonlinear growth stages of bending instability 
and show that it can be responsible for the secular increase of the stellar 
velocity dispersion in the vertical direction.

\bc
{
\bf BENDING INSTABILITY AS A MECHANISM OF THE SECULAR HEATING
OF STELLAR DISKS\\ 
}
\vspace{0.5cm}
{\it The Linear Theory}
\ec

Stellar disks are known to be unstable against the formation of spiral arms 
and bars. In addition, conditions for the growth of bending instability exist 
in thin equilibrium disks. This instability arises in systems with highly 
anisotropic particle motions and is similar to the fire-hose instability in 
plasma.

\smallskip
The bending instability of an infinitely thin gravitating layer with a 
nonzero velocity dispersion was first investigated by Toomre (1966). Toomre 
is believed to have been sceptical about his results. Therefore, they were 
published in a barely accessible paper and only several years ago did they 
become widely known after their detailed presentation by Merrit and Sellwood 
(1994). The bending instability of flat stellar systems has been rediscovered 
several times. Kulsrud~et~al.~(1971) independently obtained a result similar 
to that of Toomre (1966). As regards equilibrium cold disks (with a zero 
velocity dispersion), Hunter and Toomre~(1969) showed that they (in contrast 
to hot disks) are stable against the growth of bending perturbations. This 
fact had long been misused as an argument against any astrophysical 
applications related to bending instability. Only in 1977 did Polyachenko 
and Shukhman construct an exact linear theory for a homogeneous thin layer 
with a nonzero stellar velocity dispersion. A more realistic model of a 
thin layer with the vertical density profile described by the function 
${\rm sech}^2(z/z_0)$ was analyzed by Araki (1985)\footnote{The presentation 
of the results of this analysis by Sellwood and Merritt (1994) made it 
accessible for a wide circle of researchers.}.

\smallskip
Toomre (1966) was the first to derive the dispersion relation for 
long-wavelength ($\lambda = 2 \pi / k >> h$, where $2h$ is the layer 
thickness) bending perturbations
\be
\omega^2 = 2 \pi G \Sigma |k| - \sigma_x^2 k^2 \, ,
\label{dispersion1}
\ee
where  $\Sigma$ is the surface density of the layer stars and $\sigma_x^2$ is 
the velocity dispersion along a particular coordinate in the layer plane. It 
follows from Eq. (\ref{dispersion1}) that perturbations with wavelengths 
$\lambda > \lambda_{\rm J} \equiv \sigma_x^2 / G \Sigma$ are stable, because 
$\omega^2 > 0$ in this range.

\smallskip
Short-wavelength perturbations
$$
\lambda < \lambda_2 \approx h \frac{\sigma_x}{\sigma_z} =
h \alpha \, ,
$$
where
\be
\alpha \equiv \frac{\sigma_x}{\sigma_z} \, ,
\label{alfa}
\ee
is the control parameter of the bending instability, must also be stable. In 
this case, the time of one vertical stellar oscillation, 
$t_{\bot} = h/\sigma_z$, is longer than the time it takes for the star to 
traverse one wavelength, $t_{\|} = \lambda/\sigma_x$. Therefore, the star 
traverses a distance of several wavelengths in the time $t_{\bot}$. As a 
result, a mismatch in the coherent particle motion arises, causing the 
perturbation to decay. Intermediate-wavelength 
($\lambda_2 < \lambda < \lambda_{\rm J}$) perturbations are unstable; the 
bending only grows.

When $\lambda_2 = \lambda_{\rm J}$, the instability region disappears and 
the disk becomes stable against bending perturbations of any wavelengths. 
The following analytical estimate is valid for the parameter $\alpha$ 
defined by formula (\ref{alfa}) in the linear approximation 
(Toomre 1966; Kulsrus et al. 1971; Polyachenko and Shukhman 1977; Araki 1985):
$$
\lambda_2 = \lambda_{\rm J}\,\,\,
\mbox{\rm for}\,\,\,
\alpha = \alpha_{\rm cr} \approx 3.0\,,
$$
or
\be
\left(\frac{\sigma_z}{\sigma_x}\right)_{\rm cr} \approx 0.3 \, .
\label{crit1}
\ee
The instability is completely suppressed for $\sigma_z / \sigma_x > 0.3$ 
and grows for $\sigma_z / \sigma_x < 0.3$.

\smallskip
Stars are born out of a gaseous medium and initially have low random 
velocities. The increase of the velocity dispersion in the radial and 
azimuthal directions through scattering by spiral density waves can lead 
to an anisotropy of the particle motions in the plane and in the vertical 
direction. As a result, bending instability can develop in the disk; this 
instability causes an increase in $\sigma_z$ to the level corresponding to 
instability saturation and an increase in the stellardisk thickness. The 
linear criterion gives a low instability saturation level~(\ref{crit1}). 
However, as was mentioned above, $\sigma_z / \sigma_R \approx 0.5$ in the 
solar neighborhood. That is why Toomre was sceptical about his discovery. 
However, the following two things should be remembered. First, 
$(\sigma_z / \sigma_R)_{\rm cr} \approx 0.3$ was obtained from a linear 
analysis. Numerical studies of the nonlinear growth stages of bending 
instability (Raha et al. 1991; Sellwood and Merritt 1994; Merrit and 
Sellwood 1994; and our numerical results) show that the instability is 
saturated at much larger $\sigma_z / \sigma_R$. Second, the growth rate 
of the instability is very low. The saturation time scales are several 
billion years. 

\smallskip
For these two reasons, bending instability can play an important role in 
the secular disk heating in the $z$ direction.

\bc
{\it The Numerical Model}
\ec

A high spatial resolution in the $z$ direction is required to properly 
simulate the growth of bending instability in numerical simulations. In 
the $N$-body problem, this is possible only for a large number of 
gravitationally interacting particles. In the first numerical models, $N$ 
was taken to be 100\,000 (Raha et al. 1991; Sellwood and Merritt 1994; 
Merrit and Sellwood 1994). We used a considerably larger number of 
particles, $N = 300\,000-500\,000$, which made it possible to reach a 
higher resolution and to trace the evolution of the stellar disk on time 
scales of about 5 Gyr. In addition, our model of a disk galaxy is much 
more realistic: we considered a rotating disk with an exponential density 
profile and assumed the existence of an additional spherical component 
(a dark halo). Finally, we scanned in detail the space of control parameters 
(see the next section) and found many of the growth features of bending 
instability that were overlooked both in the linear analysis and in the 
numerical simulations.

\smallskip
In our numerical simulations, we simulated the evolution of an isolated disk 
galaxy by using an algorithm that is based on the data structuring in the 
form of an hierarchical tree (Barnes and Hut 1986). Its implementation was 
taken from the NEMO package (http://astro.udm.edu/nemo; Teuben 1995). The 
package was adapted to personal computers and significantly expanded by 
including original data analysis and visualization programs. 

\smallskip
In specifying a galaxy model, we separated two subsystems: the stellar disk 
and the spherically symmetric component (a dark halo). Star formation was 
disregarded. The disk was represented as a system of gravitating bodies with 
$R$ and $z$ density profiles, which corresponds to the observed brightness 
profiles for spiral galaxies:
\be
\label{eq_star_disk_dens}
\rho_{\rm d}(R,z)=\frac{M_{\rm d}}{4 \pi h^2 z_0}
                     \cdot \exp\left(-\frac{R}{h}\right)
                     \cdot {\rm sech}^2\left( \frac{z}{z_0} \right) \, ,
\ee
where $h$ is the exponential disk scale length, $z_0$ is the typical scale 
of density variation in the $z$ direction, and $M_{\rm d}$ is the total disk 
mass.

\smallskip
The spherical component was described in terms of the external static 
potential
\be
\label{eq_halo_halo}
\Phi_{\rm h}(r)=-\frac{v_{\infty}^2}{2}\ln(r^2+a_{\rm h}^2) \, ,
\ee
where $a_{\rm h}$ is the typical scale and 
$v_{\infty}$ is the velocity of a particle on a circular orbit in this potential for 
$r \to \infty$ (($v_{\infty}$ can be expressed in terms of the halo mass within a 
sphere of a given radius and the parameter $a_{\rm h}$).

\smallskip
We specified the initial particle velocities (the rotation velocity and the 
random component) on the basis of equilibrium Jeans equations using a 
standard technique with a specified dependence $\sigma_R^2 \propto \Sigma(R)$, 
where $\Sigma(R)$ is the disk star surface density (see, e.g., Hernquist 1993).

\begin{table}[ht]
\caption{Initial parameters of the numerical models}
\label{tab_first}
\bc
\begin{tabular}{|c|c||p{2cm}|p{2cm}|p{2cm}|}
\hline
& & \multicolumn{3}{|c|}{$Q$} \\
\cline{3-5}
$M_{\rm halo}(4h)/M_{\rm disk}(4h)$ & $z_0$ (kpc) &
\multicolumn{1}{c|}{\bf 1.5} &
\multicolumn{1}{c|}{\bf 2.0} &
\multicolumn{1}{c|}{\bf 2.2} \\
\hline
\bf 3.0 & \bf 0.1 & 9\_1& 12& \\
\hline
\bf 1.5 & \bf 0.1 & 24 & 27\_1 & 33 \\
\hline
\bf 0.85 & \bf 0.1 & 22 & 23 & 31 \\
\cline{2-5}
& \bf 0.25 & 28 & 29, 29\_1 & \\
\hline
\bf 0.6 & \bf 0.1 & 8, 8\_1 & 11, 11\_1, 11\_2, 11\_3 & 20 \\
\cline{2-5} & \bf 0.25 & 21 & & \\
\cline{2-5} & \bf 0.3 & & 26, 26\_1 & \\
\cline{2-5} & \bf 0.4 & 30 &  & \\
\cline{2-5} & \bf 0.5 & & 32 & \\
\hline
\end{tabular}
\ec
\end{table}

\smallskip
All of the parameters specified in a numerical simulation can be divided into 
three groups: input parameters of the algorithm for solving the $N$-body 
problem, parameters of the initial model, and control parameters of the 
problem (the parameters that significantly affect the processes under 
consideration).

\medskip
(1) Parameters of the algorithm: $\delta t=0.5 \times 10^6$ yr (occasionally, 
$0.25 \times 10^6$ yr) is the integration step; $T_{\rm end}=3000-5000$~Myr 
is the total integration time, $\theta=0.7$ is the parameter responsible 
for the accuracy of calculating the force (see, e.g., Hernquist 1987); 
in addition, in our computations, we took not one but the first two terms 
in the Laplace expansion of the potential: the monopole and quadrupole 
terms (Hernquist 1987); $eps=0.02$ kpc is the potential smoothing parameter.

\medskip
(2) Parameters of the initial model: $h = 3.5$ kpc is the exponential disk 
scale length; $a_{\rm h} = 2$ kpc is the dark-halo scale parameter; and 
$M_{\rm d}=(4 - 8) \times 10^{10} M_{\odot}$ is the disk mass.

\medskip
(3) Control parameters of the problem. Let us consider them in more detail.

\medskip
(A) $z_0 = 0.1 - 0.5$ kpc is the initial disk halfthickness (this quantity 
can be related to the initial velocity dispersion $\sigma_{\rm z}$ by assuming 
that the system is vertically isothermal, 
$\sigma_{\rm z}^2 = \pi G z_0 \Sigma(R)$). Since we are concerned with the 
increase of the stellar velocity dispersion in the $z$ direction, we 
constructed an initially thin equilibrium galaxy with a small value of 
$z_0$. This, in turn, implies a small velocity dispersion in the $z$ direction. 
Next, we observed the development of bending instability, which leads to 
stellar relaxation. By varying $z_0$, we were able to determine the
instability saturation level independent of the initial conditions.

\smallskip
The assumption that the system is isothermal and the dependence 
$\sigma^2_{\rm R} \propto \Sigma(R)$ taken from empirical considerations 
yields the relation $\sigma_{\rm z} / \sigma_{\rm R} = const$ for the 
initial time. The ratio $\sigma_{\rm z} / \sigma_{\rm R}$ has always been
lower than the value that follows from the linear criterion for bending 
instability. As a result, we were able to subsequently analyze the 
instability saturation level for various $R$.

\medskip
(B) $Q_{8.5} = 1.5 - 2.2$ is the Toomre parameter (Toomre 1964) at 
$R_{\rm ref}=8.5$ kpc (this quantity characterizes the initial radial 
velocity dispersion at this radius $R_{\rm ref}$). It follows from 
the dependence $\sigma^2_{\rm R} \propto \Sigma(R)$ that 
$$Q = \sigma_{\rm R} / \sigma^{\rm cr}_{\rm R} \propto
\Sigma^{1/2} \frac{\kappa}{\Sigma} \propto \kappa(R) \exp(R/2h) \, ,
$$
where $\kappa$ is the epicyclic frequency; i.e., in contrast to 
$\sigma_{\rm z}/\sigma_{\rm R}$, $Q$ is not constant for the entire disk. 
The function $Q(R)$ for an exponential disk has a broad maximum in the 
range $h < R < 3h$ (see Fig. 1 in Hernquist 1993). Specifying 
$R_{\rm ref} \approx 2.5h \approx 8.5$ kpc gives the condition 
$Q(R) \geq Q_{8.5}$. This condition, in turn, ensures a level of 
stability against perturbations in the disk plane that is not lower than 
the level specified at $R_{\rm ref}$.

\smallskip
Here, we were concerned with two cases:

\medskip
--- a large value of $Q_{8.5}$ ($\approx 2.0$) or, in other words,
a large initial radial velocity dispersion at which the formation of a 
bar in the disk is suppressed; this allows the development of bending 
instability to be traced in pure form (without the influence of a bar).

\medskip
--- a moderate value of $Q_{8.5}$ ($\approx 1.5$); in this case, the
disk is unstable against the growth of a bar mode and we were able to 
investigate the influence of a bar on the relaxation in the $z$ direction.

\medskip
(C) $M_{\rm h}(4h)/M_{\rm d}(4h) = 0.6 - 3.0$ is the ratio of
dark-halo mass to disk mass within a sphere of radius $4h$. The spherical 
component is a stabilizing factor during the development of bending 
instability; we consider its influence on the secular galaxy heating
in the section entitled ``Stellar Relaxation in Models...''.

\medskip
For our computations, we used PC-compatible computers of the Astronomical 
Institute of the St. Petersburg State University. Some of the numerical
simulations were carried out in cluster mode. The table gives the numbers 
of all of the models analyzed below\footnote{The models with equal control 
parameters have di.erent random realizations of the initial 
conditions~---~particle positions and random velocities.}.

\bc
{\it Results of the Numerical Simulations}
\ec

We carried out a large series of stellar-dynamics simulations and studied 
the growth of bending instability in thin stellar disks as a function of 
the control parameters of the problem. Analysis of the results of our 
computations revealed three distinct mechanisms of stellar relaxation in 
the vertical direction. The following system of units is used in all of 
the figures that illustrate our conclusions: the unit of time is 1 Myr;
the unit of velocity is 978 km s$_{-1}$; and the unit of length is 1 kpc.

\smallskip
{\bf Large-Scale Bending Instability of the Disk.} The large-scale bending 
instability of the entire disk is the first galaxy heating mechanism. It 
is most typical of galaxies with low-mass spherical components that are 
initially hot in the plane, i.e., of galaxies in which the formation of a 
bar was suppressed (for bar-mode suppression mechanisms, see, e.g., 
the monograph of Binney and Tremaine 1987).

\smallskip
Below, we describe the scenario for the development of bending instability 
using Model 26\_1 as an example. All of the key evolutionary features of the
stellar disk that were shown by this model were also observed in other hot 
models.

\smallskip
If we decompose $\overline{z(R)}$ (the mean particle deviation from the 
$z=0$ plane) into Fourier harmonics, 
${\bar z}_{m}(R) = A_{m} \exp{(-im \varphi)}$, then we can calculate
the amplitudes of the first three harmonics ($m = 0$ is the axisymmetric 
bending or, alternatively, the bell mode; $m = 1$ is the bending mode; 
and $m = 2$ is the saddle-type mode). The change of $A_m$ with time describes 
the various bending formation stages shown in Figs. 1 and 2 (left panels). 
These stages are more clearly distinguished in the color two-dimensional 
histograms that can be found at 
http://www.astro.spbu.ru/staff/seger/articles/warps\_2002/fig1\_web.html. 
In these histograms, the mean disk particle deviation from the $z = 0$ 
plane is represented by different colors (the shades of yellow and blue 
indicate upward and downward deviations, respectively). We see in 
Figs. 1 and 2 (as in the two-dimensional histograms) that a 
large-scale bending is produced in the galaxy at a time 
$t\approx 200$ Myr. At $t\approx 400$ Myr, the bending amplitude reaches 
its local maximum. The bending perturbation wavelength (the radial extent) 
is comparable to the disk scale length. The bending is not axisymmetric; 
the amplitude of the zero mode, i.e., the mode with the azimuthal number
$m = 0$, is small, at least less than the amplitudes of other harmonics. 
This relationship between the harmonic amplitudes is preserved until 
$t\approx 800$ Myr.

\smallskip
Next, a steadily growing axisymmetric bending mode ($m = 0$) is clearly 
revealed, while all of the still large (in amplitude) nonaxisymmetric 
modes ($m = 1$ and $m = 2$) are displaced to the galactic periphery.
Subsequently ($t\approx 800-1200$ Myr), the nonaxisymmetric bending modes 
are gradually damped, while the axisymmetric bending reaches its maximum 
($t\approx 1000$ Myr).

\smallskip
For the subsequent 1500 Myr of its evolution, the disk freezes in a shape 
similar to the shape of circles on water. A similar vertical disk structure, 
with box-shaped iosphotes in central regions, was also observed in the 
numerical simulations of Sellwood and Merritt (1994) for initially hot 
Kuzmin~--~Toomre disk models and in other simulations in which the bar 
formation was suppressed through a large Toomre parameter $Q$ 
(Patsis et al. 2002). In our computations, the amplitude of the axisymmetric 
bending slowly decreased and, to all appearances, the bending must 
subsequently disappear altogether.

\smallskip
Two evolutionary stages of the bending can be separated in this model: 
the initial stage with a nonaxisymmetric bending and the main stage with an
axisymmetric bending.

\smallskip
Let us turn to the plot of the vertical ($\sigma_z$) and radial ($\sigma_R$) 
velocity dispersions against time (Fig. 3a). The value of $\sigma_z$ 
approximately doubled in time $t = 3000$ Myr. Two times of rapid increase 
in the velocity dispersion $\sigma_z$ can be distinguished: one at 
$t\approx 500$~Myr (the time the amplitude of the initial nonaxisymmetric 
bending reaches its maximum) and the other at $t\approx 1000 - 1200$~Myr 
(the time the amplitude of the main axisymmetric bending reaches its
maximum). As we see, the two times of change in $\sigma_z$ are closely 
related to the characteristic growth stages of bending instability. Such 
a relationship exists in all of the models without bars in which bending
instability developed. Thus, we can conclude that in this model, the 
bending instability is responsible for the increase in $\sigma_z$; when 
the bell ($m = 0$) mode appears, the energy of the random stellar velocity in
the disk plane is converted most effectively into the energy of random 
vertical motions.

\smallskip
Note that the integrated quantities $\sigma_z$ and $\sigma_R$ calculated
for the entire galaxy were used to construct the plots in Fig. 3. These 
quantities allow us to judge only the general run of the processes in 
the disk. Figure 4 shows the radial profiles of the azimuthally averaged 
$\sigma_z$ for various times. We see that the relaxation effect is most 
pronounced for central regions.

\smallskip
In Fig. 5, the $\sigma_z / \sigma_R$ ratio is plotted against $R$ for 
several times (here, we also averaged $\sigma_z / \sigma_R$ in concentric 
rings). We can conclude from our data that the saturation level of the 
bending instability in a region of about two exponential disk scale lengths
in size (about 7~--~7.5 kpc) ismuch higher than its level predicted by the 
linear criterion (\ref{crit1}). For these regions, 
$\sigma_z / \sigma_R \approx 0.4 - 0.8$.

\begin{figure}
\centerline{\psfig{file=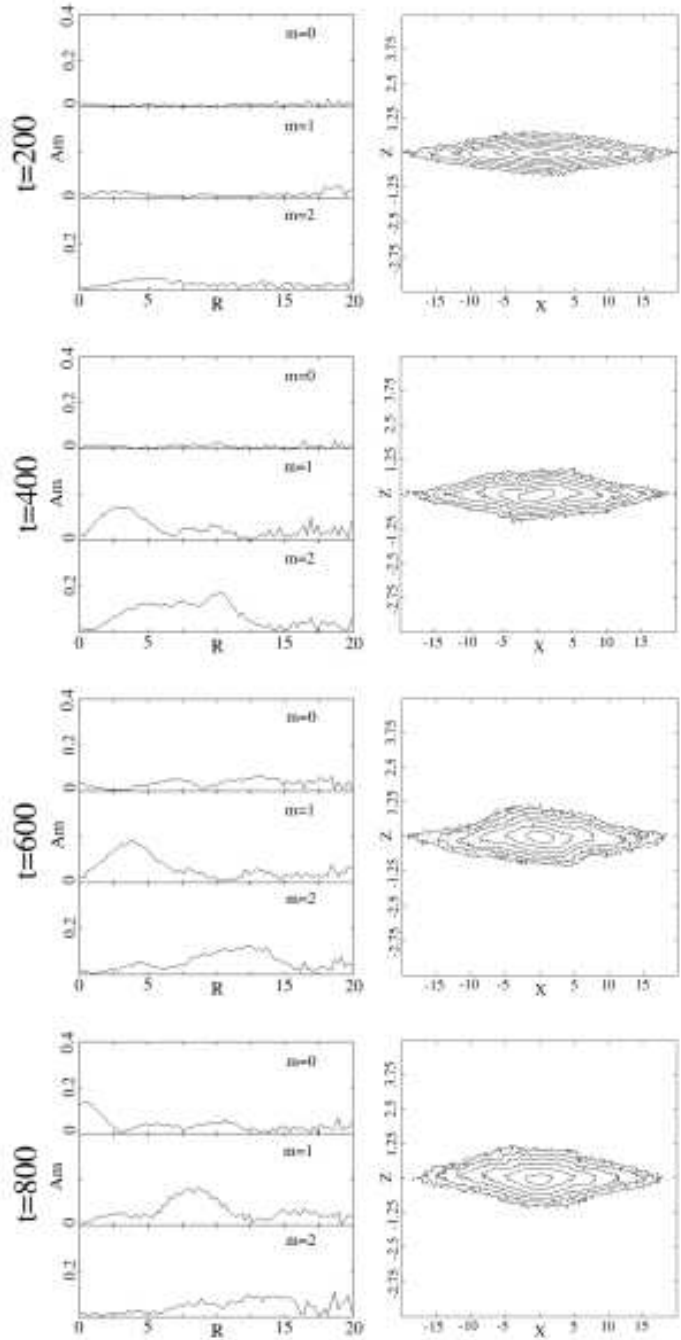,width=10cm}}
\caption[1]{\small Model 26\_1. Early evolutionary stages of the bending. The left 
frames: the radial distribution of the amplitude of the first three harmonics 
($m =0, 1, 2$) for the bending perturbation for several times. The right frames: 
an edge-on view of the galaxy --- the isophotal distribution. The horizontal 
and vertical frame sizes are 40 and 10 kpc, respectively; i.e., {\bf the vertical 
scale was increased by a factor of 4!}}
\end{figure}

\begin{figure}
\centerline{\psfig{file=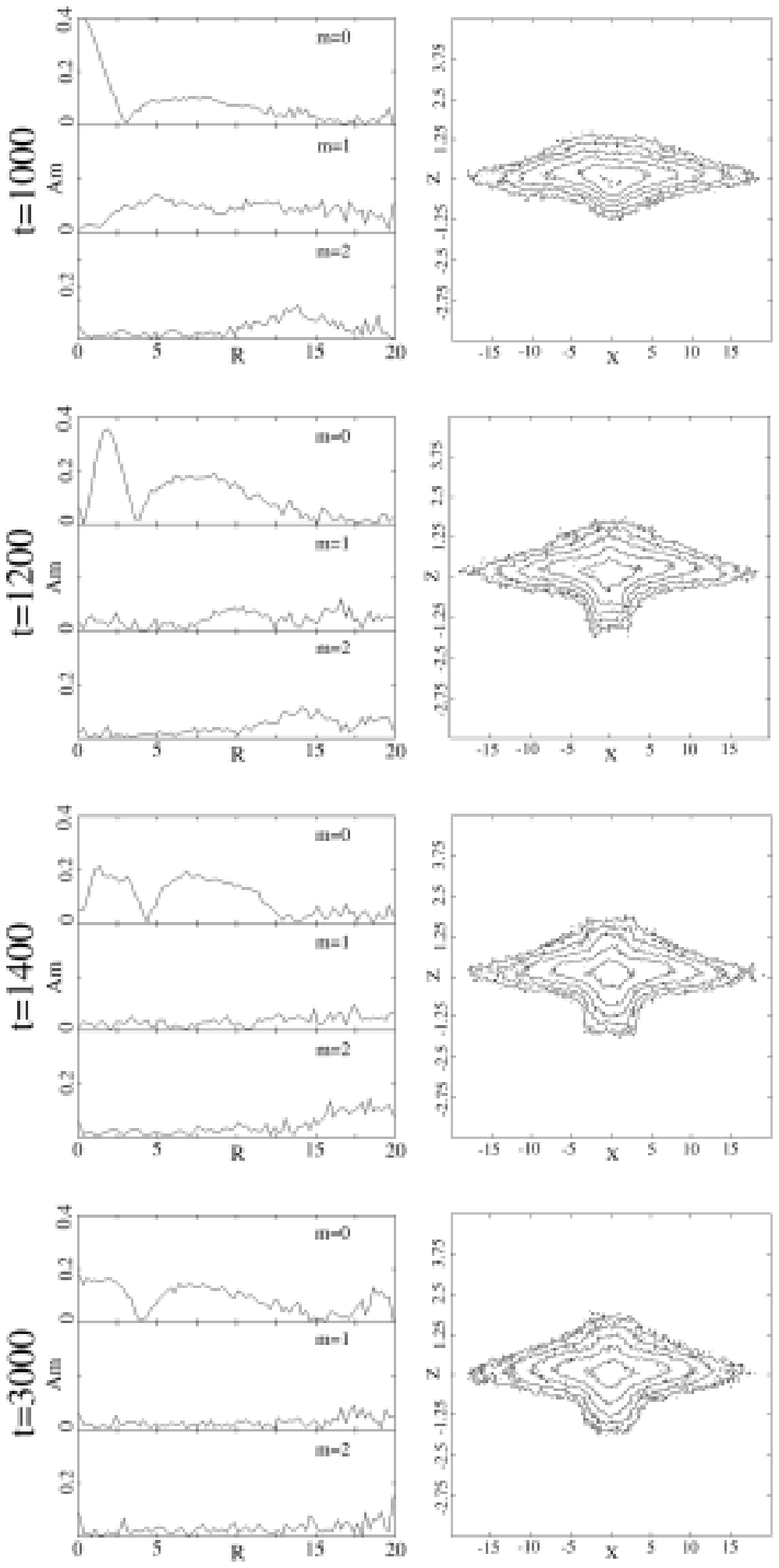,width=10cm}}
\caption[2]{\small Model 26\_1. The same as Fig. 1 for late evolutionary stages.}
\end{figure}

\begin{figure}
\centerline{\psfig{file=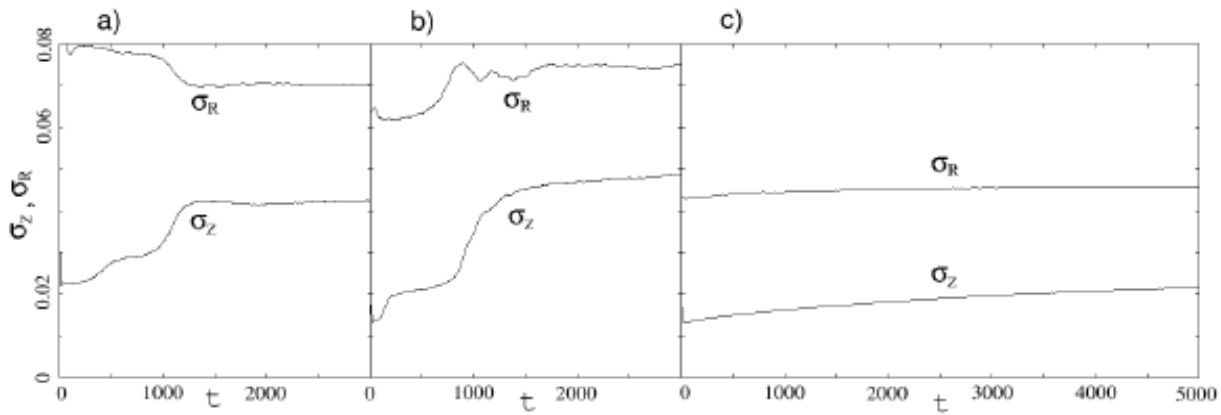,width=17cm}}
\caption[3]{\small $\sigma_R$ (upper curve) and $\sigma_z$(lower curve) versus time. 
The integrated quantities referred to the entire galaxy are plotted: 
(a) Model~26\_1; (b) Model~8\_1; and (c) Model~9\_1.}
\end{figure}

\begin{figure}
\centerline{\psfig{file=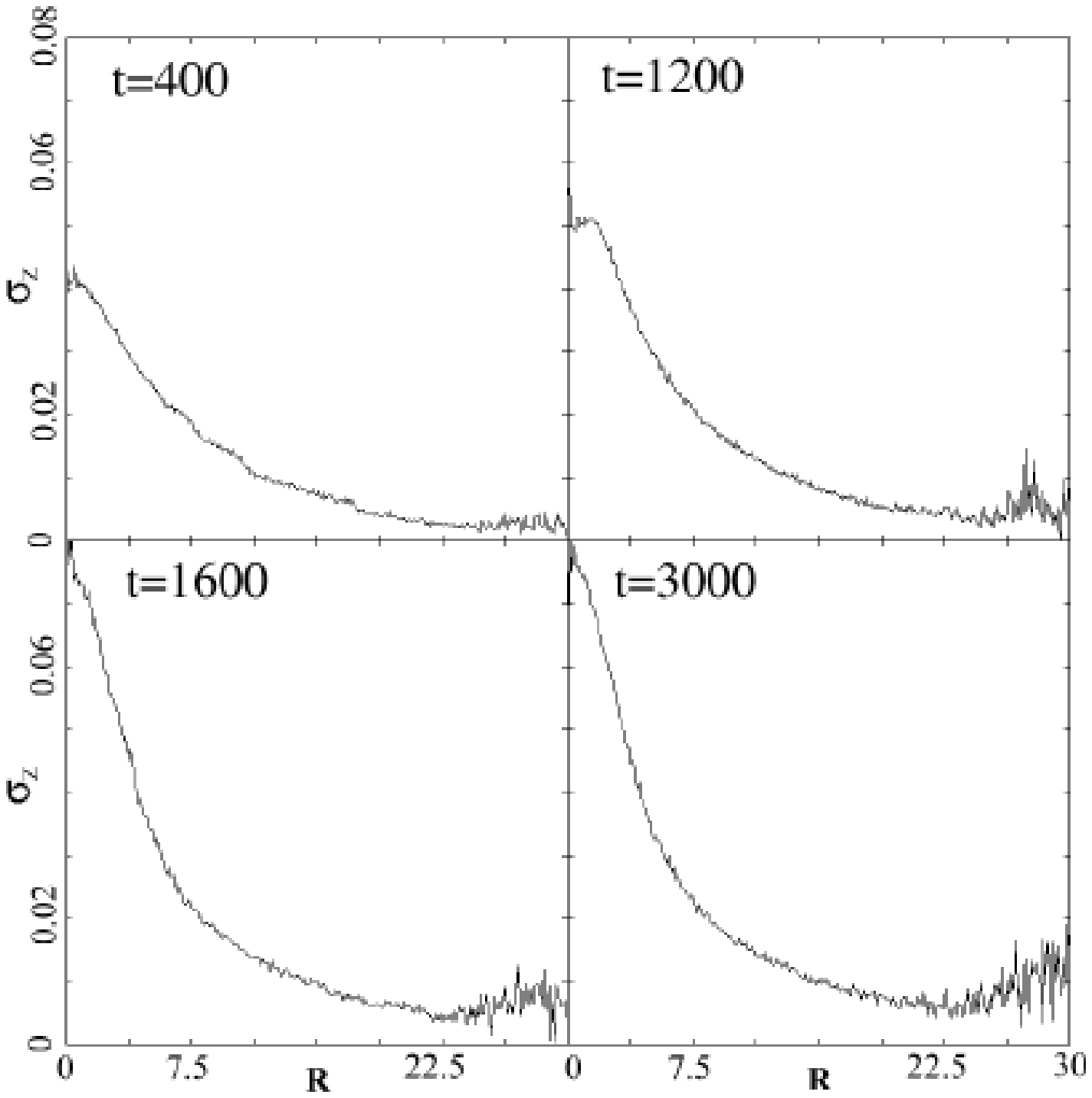,width=17cm}}
\caption[4]{\small Model~26\_1. $\sigma_z$ versus $R$ for several times.}
\end{figure}
 
\smallskip
In Figs. 1 and 2 (right panels), the edge-on image of a model galaxy is 
represented by isophotes for several times. The image was vertically magnified
by a factor of 4. As a result, the disk bending is clearly seen. Note that 
the galactic evolution at late stages gives rise to a family of X-shaped orbits in
central regions of the disk ($t = 1200, 1400, 3000$; the left frames; the 
X-shaped structures are more clearly seen on the pseudoimage of the model 
galaxy, 
http://www.astro.spbu.ru/staff/seger/articles/warps\_2002/fig1\_web.html). 
The main axisymmetric bending may be saturated at a galaxy thickness
at which a resonance arises between the frequencies of the stellar 
oscillations across the disk and in the rotation plane. This resonance 
can produce X-shaped orbits, which have repeatedly been observed in 
numerical simulations (Combes et al. 1990; Pfenninger and Friedli 1991; 
Patsis et al. 2002).

\begin{figure}
\centerline{\psfig{file=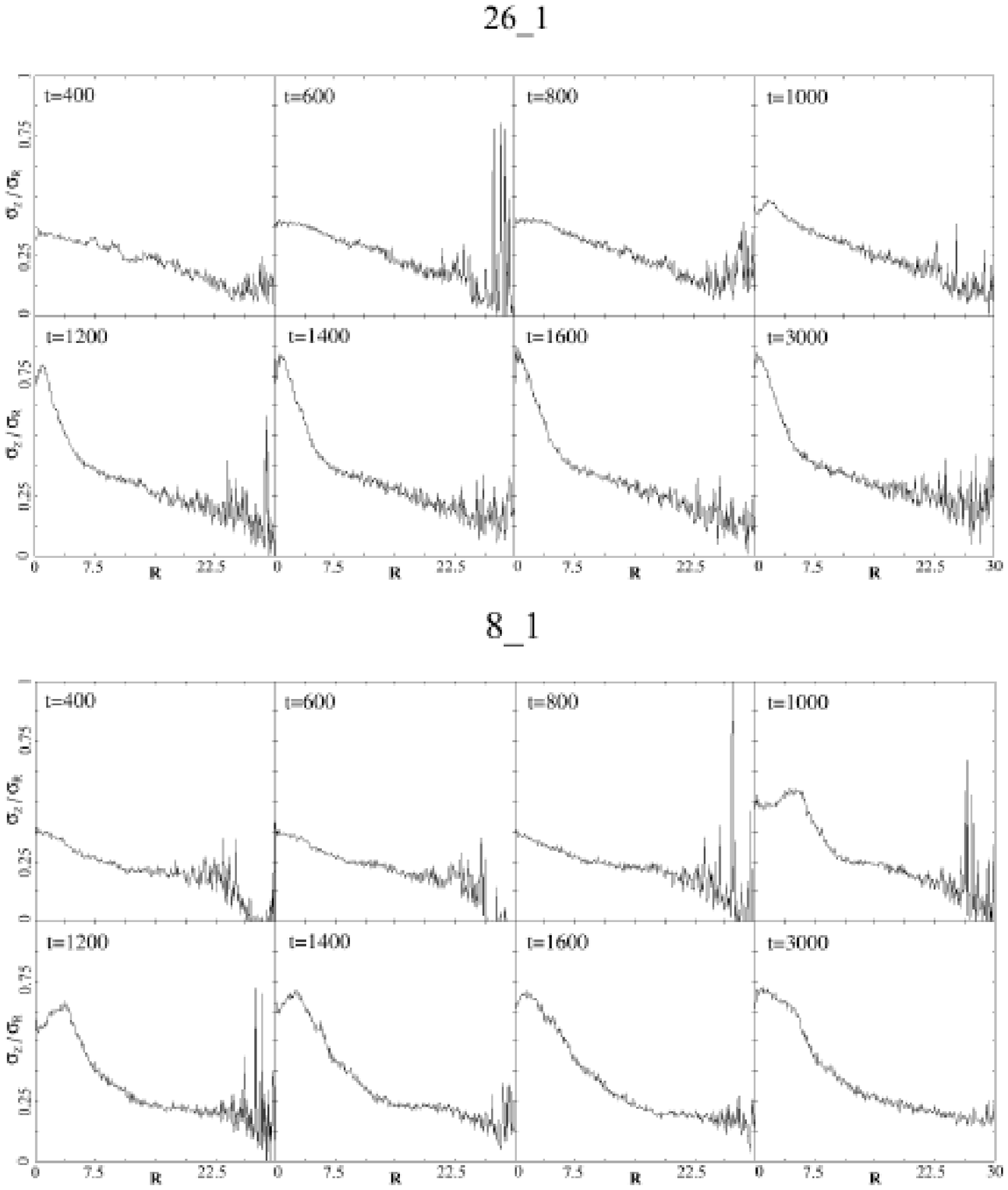,width=17cm}}
\caption[5]{\small $\sigma_z/\sigma_R$ versus $R$ for several times: Models 26\_1 and 8\_1.}
\end{figure}

In other models, the initial and main bends manifested themselves differently that 
those in Model 26\_1. The behavior of all models can be assigned to one of the 
following cases.

\medskip
(1) Both the initial and main bends arose almost simultaneously. In the 
long run, only the axisymmetric bending survived. Abrupt disk heating took 
place at the time the bending amplitude was increasing. This scenario was 
observed for initially thin galaxies, i.e., for the models that started from 
initial conditions far from the linear saturation level. Almost all of the
models with $Q_{8.5}=2.0$ and $z_0<0.3$ kpc show such a pattern of evolution.

\medskip
(2) The occurrences of the initial and main bends were well separated in time, 
as for Model 26\_1 described in detail.

\medskip
(3) There was no initial bending at all, while the main bending could be observed 
at very late evolutionary stages. This scenario is characteristic of initially 
thick disks, i.e., of the models that started from conditions close to the 
instability saturation limit. This result closely corresponds to the following
theoretical result: at a given ratio of the masses of the spherical and disk 
components, the mode with the azimuthal number $m = 1$ becomes stable earlier
than the other modes as the disk thickness increases (Fridman and Polyachenko 1984). 
As the disk thickness increases further, the m = 2 mode is ultimately saturated and 
the $m = 0$ mode remains at the fore. Note that the results of our numerical simulations
and the theoretical calculations of Fridman and Polyachenko (1984) are inconsistent 
with the conclusions of Merrit and Sellwood (1994). These authors argue that the 
$m = 0$ mode is primarily stabilized in thick (radially hot) disks and the $m = 1$ 
mode must remain at the fore. In our numerical simulations, as the disk thickened, 
the growth rate of the $m = 0$ mode decreased sharply. A low-amplitude $m = 1$ 
mode was initially observed against the background of a slowly growing $m = 0$ 
mode; after the $m = 1$ mode displaced to the periphery, an axisymmetric bending 
clearly showed up. The vertical velocity dispersion in thick models began to 
increase only after the axisymmetric bending developed. Thus, for example, the 
disk bending in Model 32 was observed only 3000 Myr after the beginning of the 
evolution.

\medskip
(4) In some of the models in which a bar was formed, only a small increase in the 
vertical velocity dispersion was associated with the initial bending of the entire 
disk. This applies to all of the models with $Q_{8.5}=1.5$ and $z_0<0.3$ kpc. 
For thicker galaxies, no initial bending of the entire disk whatsoever was 
observed. In all of the models with $Q_{8.5}=1.5$, the bending instability 
of the bar itself, which is investigated in the next subsection, was the 
leading heating mechanism.

\medskip
{\bf The Bending Instability of Bars.} The second vertical disk heating mechanism 
revealed by our numerical simulations is related to the bending instability of bars. 
A bar bending was first detected by Raha et al. in three-dimensional numerical 
simulations. Its appearance was explained by the development of firehose instability 
in the bar. In our simulations, this instability was responsible for the secular 
disk heating almost in all of the models with bars\footnote{In some of the models 
that started from $Q_{8.5}=2.0$, despite the large velocity dispersion in the disk 
plane, a distinct bar was formed at late stages. However, by this time, the vertical
velocity dispersion was so large (because of the relaxation associated with the 
bending instability of the entire disk) that no bending was formed in the bar.}.

\medskip
Consider the disk evolution scenario using Model 8\_1 as an example. The main 
evolutionary stages of the bending in this initially moderately hot model are 
shown in the color two-dimensional histograms (Fig. 6 and 7). 
At early stages ($t \approx 200-400$ Myr), the initial bending of the 
entire disk comes to the fore (the group of frames in the middle column). 
Gradually, the initial bending perturbation reaches its saturation level, 
is drifted to the galactic periphery, and decays ($t \approx 600$ Myr). A 
distinct bar has already been formed in the galaxy by this time (Figs. 6 
and 7; the group of left frames; all images were oriented in such a way that 
the bar major axis was horizontal). At $t \approx 800$ Myr, the bending is
generated already in the bar. At $t \approx 1000$ Myr, its amplitude reaches 
a maximum, after which the bending perturbation rapidly decays 
($t \approx 1200-1600$ Myr).

\medskip
The bar bending effect becomes understandable if we look at the group of right 
frames in Figs. 6 and 7. The shades of gray in these figures indicate the disk
thickness in different disk regions. The thickness was calculated as 
$\sqrt{\overline{z^2(R, \varphi)} - \overline{z(R, \varphi)}^2}$. We see that the
bar was much thinner than the rest of the galaxy, which can be explained as 
follows. Since most of the disk stars were captured into the bar, it has a high 
surface density. Self-gravity causes the disk to become thinner in the bar region, 
which gives rise to a bending.

\medskip
If we look at the plot of the vertical velocity dispersion against time (Fig. 3b), 
then we will see the curve rises twice: first at $t \approx 200$ Myr, which 
is clearly associated with the appearance of the initial bending of the entire 
disk, and, second, at $t \approx 1000-1200$ Myr, which coincides with the time the 
amplitude of the bar bending reaches its maximum.

\medskip
In Fig. 5, $\sigma_z / \sigma_R$ is plotted against $R$ for several times. We see 
that for relatively cold disks (as in the case of hot disks; Fig. 5, Model 26\_1), 
the general saturation level of the bending instability is higher than the linear 
level ((\ref{crit1}). Merrit and Sellwood (1994) pointed out that moderately hot 
models ($Q \sim 1$) behave virtually as predicted by the linear theory. Large
deviations take place only for hot models. In our simulations, we observed a 
similar picture. The saturation level of the bending instability associated with
the entire disk in the models with $Q_{8.5}=1.5$ is close to the linear level 
(see Fig. 5, $t = 800$, Model 8\_1). However, when the bending instability of the 
forming bar came into effect, the final vertical disk heating was almost the same 
as that for the models with $Q_{8.5}=2.0$ 
(see Fig. 5, $t = 3000$, Models 26\_1 and 8\_1).

\begin{figure}
\centerline{\psfig{file=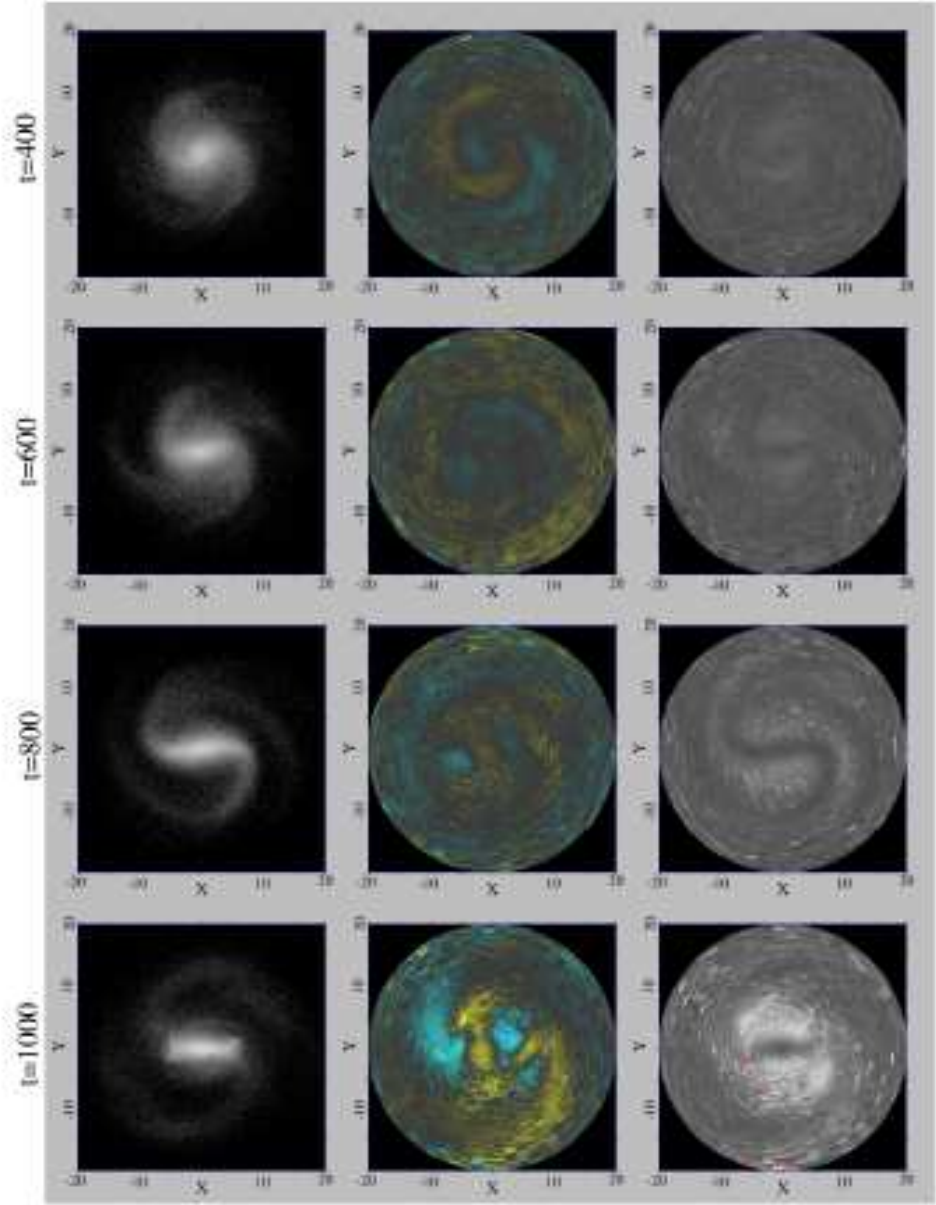,width=14cm}}
\caption[6]{\small Model 8\_1. The evolution of a disk with $Q_{8.5}=1.5$ early stages). 
In each row: 
the first frame shows the galaxy seen face-on (the image brightness corresponds 
to the logarithm of the number of particles per pixel); 
the second frame shows the mean disk particle deviation from the $z = 0$ plane 
(this value is represented by different colors: the shades of yellow and blue 
indicate upward and downward deviations, respectively; the deviation ranges 
from $-0.5$ to $+0.5$ kpc);
the third frame shows the two-dimensional galaxy thickness distribution (the 
thickness is indicated by the shades of gray: the lighter is the disk region, 
the thicker it is; the thickness ranges from 0 to 1 kpc; the red point means 
that the thickness in this region is outside the range). 
All pictures are oriented in such a way that the bar was located 
along the $X$ axis. The frame size is 40 kpc $\times$ 40 kpc.}
\end{figure}

\begin{figure}
\centerline{\psfig{file=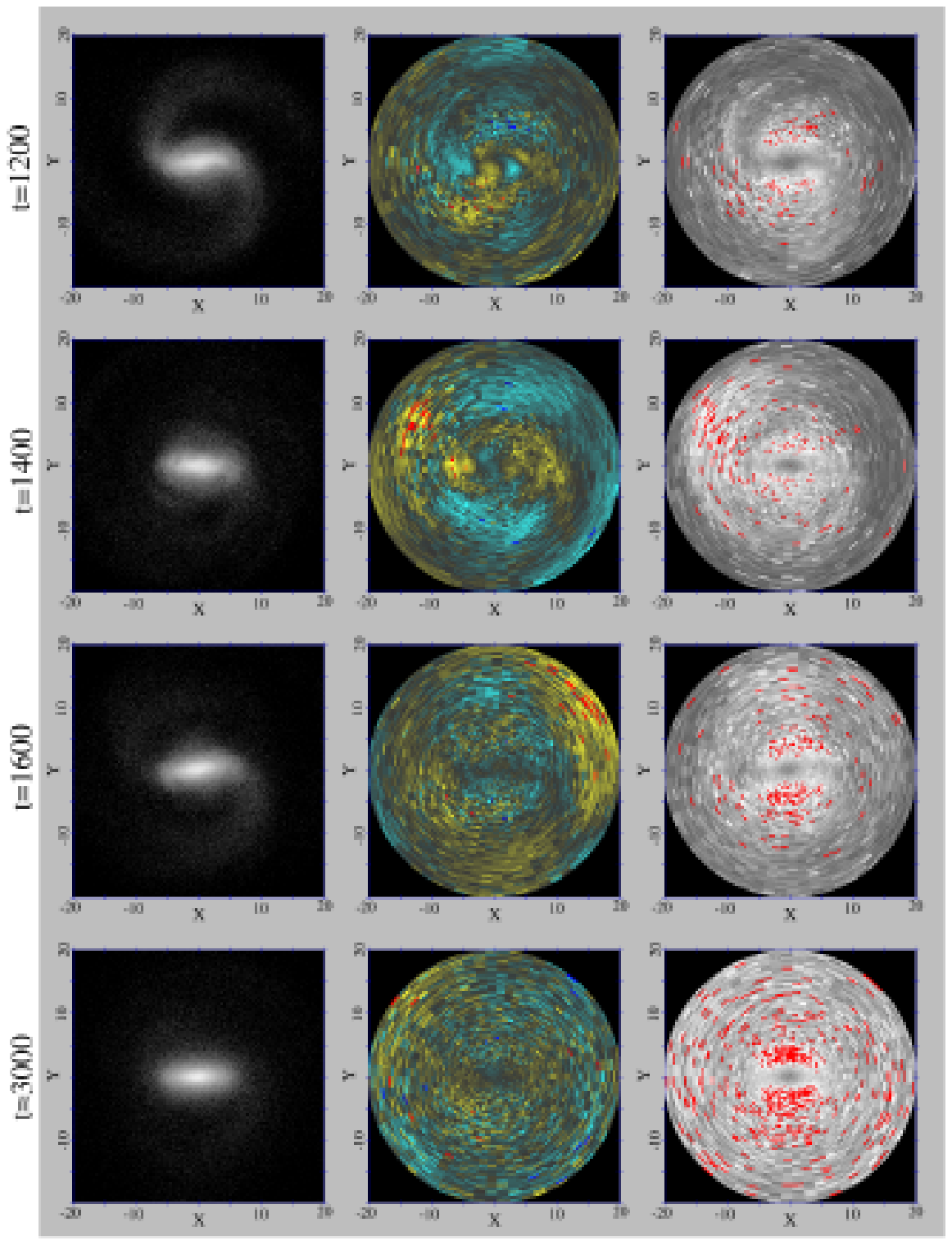,width=14cm}}
\caption[7]{\small Model 8\_1. The same as Fig. 6 for late evolutionary stages.}
\end{figure}

For the remaining models with $Q_{8.5}=1.5$ and a low dark-halo mass 
($M_{\rm h}(4h)/M_{\rm d}(4h) < 1$), we observed the same growth stages 
of bending perturbations as those in Model 8\_1. If a bar was formed
in the disk, then bending instability sooner or later began to develop 
in it. The bending amplitude rapidly increased and the instability was 
ultimately saturated (all of these processes took place on time scales of 
the order of one billion years). Note that even the bending shape was in 
each case similar to that observed in Model 8\_1. The models differed only 
by the bar formation time and by the duration of the stage that preceded 
the onset of bending formation in the bar itself. The larger was the 
dark-halo mass, the later the bar was formed and the farther was the time of
bending generation in the bar from this time. The larger was the initial 
disk thickness, the later the bar was formed.

\medskip
{\bf Stellar Relaxation in Models with a Massive Halo.} Consider the stellar 
relaxation for models with a massive halo. A massive spherical component is
known to effectively suppress the growth of bar-like instability 
(Ostriker and Peebles 1973) and to have a stabilizing effect on the growth of 
bending perturbations (Zasov et al. 1991; Sellwood 1996). The stellar relaxation 
in such models (if it takes place) must be produced by some other factors. 
Let us trace the evolution of the vertical structure of a stellar disk embedded 
in a massive halo using Model 9\_1 as an example.

\smallskip
Because of the presence of a massive spheroidal component in the galaxy, 
the bending instability of its disk was suppressed. No bar was formed either 
(at least on a time scale of 5 Gyr). Since a massive halo primarily stabilizes 
the disk against the growth of perturbations in the disk plane with 
azimuthal numbers $m \leq 2$, higher-order modes come to the fore. They
manifest themselves in the form of multiple shortlived spiral waves and 
persist in the disk for several disk rotations 
(see, e.g., the left frame in Fig. 8). Their amplitude is initially large 
but the spiral pattern is blurred almost completely by the time $t = 3000$ Myr.

\begin{figure}
\centerline{\psfig{file=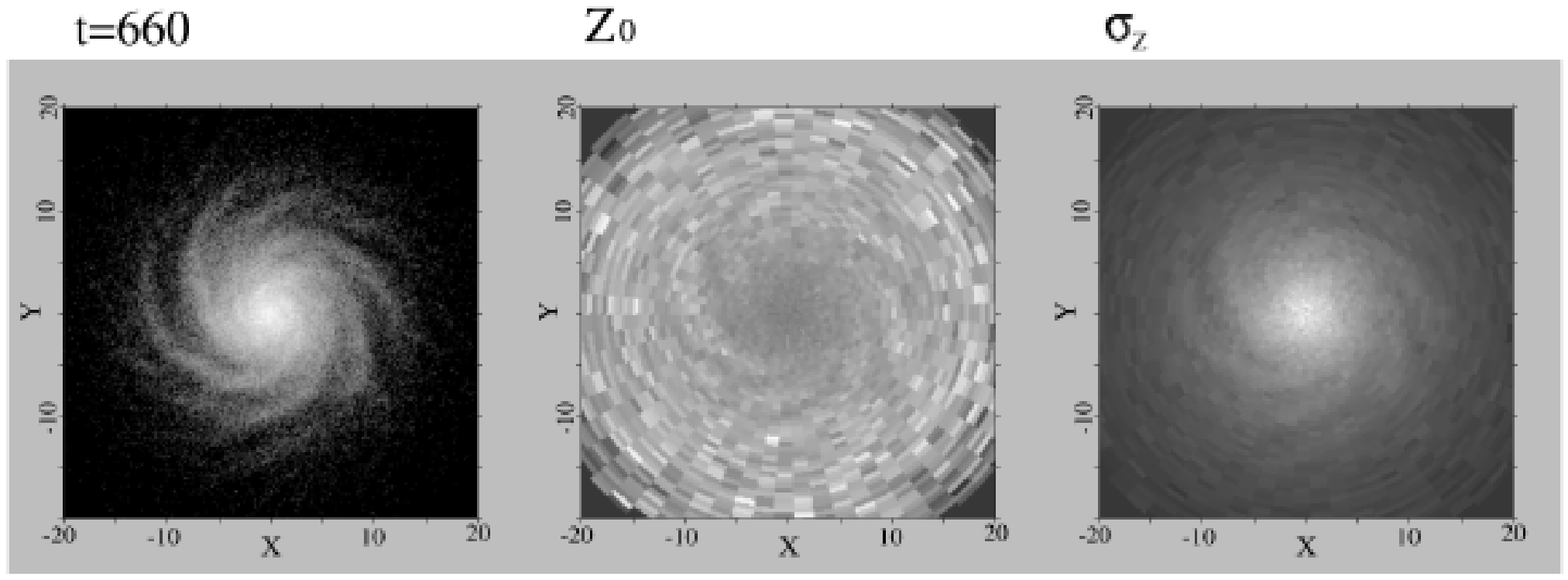,width=17cm}}
\caption[8]{\small Model 9\_1. A snapshot of the galaxy at $t = 660$ and the two-dimensional 
disk thickness and $\sigma_z$ distributions. The size of each frame is 
40 $\times$ 40 kpc. The left frame shows the galaxy seen face-on (the image brightness 
corresponds to the logarithm of the number of particles per pixel). The middle frame 
shows the two-dimensional disk thickness distribution constructed by using the shades 
of gray (the brighter the disk region, the larger its thickness at a given location). 
The right frame shows the two-dimensional $\sigma_z$ distribution over the galaxy; 
$\sigma_z$ is indicates by the shades of gray (the velocity dispersion is larger in brighter 
regions).We see that in the region of the spiral arms, the galaxy is thinner but the 
vertical velocity dispersion is larger. This effect is most pronounced for the two 
arms located in the lower right corner.}
\end{figure}

Transient spirals are produced by collective processes and they are responsible 
for the heating of the disk in its plane. This relationship was first pointed
out by Sellwood and Carlberg (1984). In Fig. 3c, the radial ($\sigma_R$) and 
vertical ($\sigma_z$) velocity dispersions are plotted against time. As would be 
expected, the rate of increase in the radial velocity dispersion slowly decreases
with decreasing amplitude of the spiral arms and, after $t = 1500$ Myr, the 
stellar relaxation in the disk plane related to transient spiral perturbations
becomes ineffective. At the same time, for all five billion years of evolution, 
$\sigma_z$ slowly increases, although the rate of increase gradually decreases. 
The observed vertical secular disk heating is not the result of numerical pair 
relaxation, because the levels and the patterns of increase of $\sigma_z$ 
and $\sigma_R$ differ greatly ($\sigma_R$ is almost constant at late 
evolutionary stages). The cause of the increase of $\sigma_z$ in our 
simulations is completely different.

\medskip
Let us analyze the vertical disk structure in more detail. Figure 8 shows 
three frames for one of the times: the galaxy seen face-on, the two-dimensional
disk thickness map, and the two-dimensional vertical velocity dispersion map. 
We clearly see from these frames that the disk thickness in the regions where
the spiral arms are located is smaller than the disk thickness in the 
interarm space. In other words, inhomogeneities in the distribution of 
stars in the plane produce inhomogeneities in the vertical distribution of
stars. The effect is similar in nature to the bar thinning described in the 
preceding section: a self-gravitating disk is thinner where the surface 
density is higher.

\medskip
We assume that the observed increase of the vertical velocity dispersion 
is related to the scattering of stars by inhomogeneities in the distribution 
of matter in the $z$ direction. Apart from the above explanation, there are 
two more additional facts that are indicative of this.

\medskip
If we look at the two-dimensional $\sigma_z$ distribution (Fig. 8), then 
we can see that $\sigma_z$ in the spiral arms is higher than that in the 
interarm space. In addition, it follows from our computations that the 
decrease in the rate of increase of $\sigma_z$ well correlates with the
decrease in the intensity of transient spirals.

\medskip
Thus, we believe that the slow increase of the vertical velocity dispersion 
in our described models is attributable to the scattering of stars by 
transient spiral perturbations. The latter produce inhomogeneities in
the vertical distribution of matter due to self-gravity.

\medskip
{\bf Vertical Secular Disk Heating Mechanisms.} During our numerical 
stellar-dynamics simulations, we scanned the space of control parameters 
(the initial disk half-thickness, the degree of disk heating in the disk 
plane, and the relative mass of the spheroidal subsystem). A detailed 
analysis of our numerical results revealed three distinct mechanisms 
of secular disk heating in the z direction.

\smallskip
(1) The large-scale bending instability of the entire disk.

\smallskip
(2) The bending instability of bars.

\smallskip
(3) The heating due to inhomogeneities in the vertical distribution of stars 
produced by matter inhomogeneities in the plane.

\smallskip
The action of a particular heating mechanism depends on the control parameters. 
If the bar mode is suppressed in the galaxy, then the heating is attributable
to the large-scale bending instability associated with the entire disk. This is 
the case of a hot disk ($Q_{8.5} > 1.5$) and (or) a moderately massive halo. 
In this case, the heating of the central regions is particularly strong. 
The saturation level of the bending instability in these regions is almost a
factor of 2.5 higher than the level that follows from the linear theory. 
The bending mode with the azimuthal number $m = 0$ plays a key role in 
the secular heating. The time of its occurrence depends on the initial disk
thickness: the thicker is the disk, the farther in time is the secondary rise 
in disk temperature related to the growth of the bell mode.

\smallskip
If a bar mode develops in the disk, then the bending instability of the bar 
gives the largest contribution to the heating. Until a bar is formed,
the maximum
value of $\sigma_z/\sigma_R$ is 0.37 in the central regions and 0.3 on the 
periphery. This is the level that the bending instability associated with 
the entire disk reaches. It is in good agreement with the linear 
criterion. After the formation of a bar, at the time of its bending in the
central regions, $\sigma_z/\sigma_R$ rises to 0.7~--~0.8.
The heating on vertical 
bar bending perturbations dominates in initially moderately hot disks 
($Q_{8.5} = 1.5$) with a low-mass halo.

\smallskip
The heating on vertical inhomogeneities take place when the bending modes 
and the bar mode were suppressed. This was observed for moderately
hot models ($Q_{8.5} = 1.5$) with a massive halo. Note that the secular 
disk heating in all directions in similar simulations has been observed 
more than once. However, we are probably the first to associate the 
increase of $\sigma_z$ with the scattering of stars by vertical 
inhomogeneities.

\bc
{\bf CONCLUSIONS}
\ec

We have numerically analyzed the nonlinear growth stages of bending 
instability in stellar disks with exponential radial density profiles 
and found significant deviations from the linear theory.

\smallskip
(1) All of the observed modes are global; i.e., the scale of unstable 
perturbations is larger than the typical scale of density variations 
in the disk. Our conclusion agrees with the conclusions of Sellwood (1996).
It implies that, although the dispersion relation (\ref{dispersion1}) 
derived for a homogeneous layer is also locally valid for inhomogeneous 
disks (in particular, exponential disks, as in our numerical 
simulations), it would be inappropriate to use this relation to analyze 
the saturation level of long-wavelength perturbations. This also suggests 
that bending instability will develop in inhomogeneous disks differently 
in different parts of the galaxy.

\smallskip
(2) The value of $(\sigma_z / \sigma_R)_{\rm cr} \approx 0.3$ was obtained
from a linear analysis. As our numerical computations show, the saturation 
level of large-scale bending perturbations is a factor of 2 or 3 higher 
than the linear level. The differences are best seen in central regions of 
the stellar disk. The lower the dark-halo mass, the higher the instability 
saturation level and the larger the ratio of the vertical and radial velocity 
dispersions averaged within two exponential disk scale lengths. A similar 
dependence was pointed out by Mikhailova et al. (2001), although without 
discussing the underlying mechanisms. Early-type spiral galaxies have, on
average, a smaller dark-to-luminous mass ratio and, as follows from the 
observational data given above, a larger value of 
$\sigma_z / \sigma_R$, in agreement with our computations.

\smallskip
(3) The instability saturation time scales are several billion years.

\smallskip
(4) Our numerical simulations revealed three distinct mechanisms of 
the secular heating of stellar disks in the $z$ direction. We confirmed 
the existence of bar bending instability that was first detected by
Raha et al. (1991). For a large series of models, we showed that the 
bar bending is an inevitable stage of its evolution.

\smallskip
Thus, we can conclude that bending instability can play an important role 
in the secular disk heating in the $z$ direction.

\bc
ACKNOWLEDGMENTS
\ec

This work was supported in part by the Program ``Leading Scientific 
Schools'' (project no. 00-15-96607), the Federal Program ``Astronomy'' 
(project no. 40.022.1.1.1101), and the UR grant no. 02.01.006.

\bc
REFERENCES
\ec

\begin{flushright}
Translated by V. Astakhov
\end{flushright}

\end{document}